\documentclass[aps,prl,longbibliography,twocolumn,superscriptaddress]{revtex4-2}
\usepackage{amsmath}
\usepackage{textcomp}
\usepackage{hyperref}
\usepackage{graphicx}
\usepackage{float}
\usepackage{soul}
\usepackage{xcolor}

\begin{document}
\title{Generalized Gouy rotation of electron vortex beams in uniform magnetic fields}
\author{Qi Meng}
%\email{mengq8@mail2.sysu.edu.cn}
\affiliation{Sino-French Institute of Nuclear Engineering and Technology, Sun Yat-Sen University, Zhuhai 519082, China}
\author{Xuan Liu}
%\email{liux666@mail2.sysu.edu.cn}
\affiliation{Sino-French Institute of Nuclear Engineering and Technology, Sun Yat-Sen University, Zhuhai 519082, China}
\author{Wei Ma}
%\email{mawei25@mail.sysu.edu.cn}
\affiliation{Sino-French Institute of Nuclear Engineering and Technology, Sun Yat-Sen University, Zhuhai 519082, China}
\author{Zhen Yang}
%\email{yangzh97@mail.sysu.edu.cn}
\affiliation{Sino-French Institute of Nuclear Engineering and Technology, Sun Yat-Sen University, Zhuhai 519082, China}
\author{Liang Lu}
%\email{luliang3@mail.sysu.edu.cn}
\affiliation{Sino-French Institute of Nuclear Engineering and Technology, Sun Yat-Sen University, Zhuhai 519082, China}
\affiliation{United Laboratory of Frontier Radiotherapy Technology of Sun Yat-sen University \& Chinese Academy of Sciences Ion Medical Technology Co., Lid., Guangzhou 510000, China}
\author{Alexander J. Silenko}
%\email{alsilenko@mail.ru}
\affiliation{Bogoliubov Laboratory of Theoretical Physics, Joint Institute for Nuclear Research, Dubna 141980, Russia}
\author{Pengming Zhang}
%\email{zhangpm5@mail.sysu.edu.cn}
\affiliation{School of Physics and Astronomy, Sun Yat-sen University, Zhuhai 519082, China}
\author{Liping Zou}
\email[Corresponding author.\\]{zoulp5@mail.sysu.edu.cn}
\affiliation{Sino-French Institute of Nuclear Engineering and Technology, Sun Yat-Sen University, Zhuhai 519082, China}

\begin{abstract}
The intrinsic rotation of electron vortex beams, governed by their phase structure, has been experimentally observed in magnetic fields by breaking the beam's cylindrical symmetry. However, conventional Landau states, which predict three fixed angular frequencies, cannot fully account for the existing experimental observations. To address this limitation, we introduce and derive the generalized Gouy rotation angle, which links the Gouy phase of an extended Landau state---featuring a periodically oscillating beam width---to the experimentally observed angular variation. In particular, this framework predicts a broader spectrum of angular frequencies and captures the reversal of rotation direction observed in electron vortex beams with negative topological charge. Calculations based on experimental parameters show good agreement with previously published data and are further validated here by numerical simulations using the Chebyshev method. Our results are, in principle, applicable to any system involving electron vortex beams in uniform magnetic fields, and provide a foundation for exploring vortex electrons in Glaser and other nonuniform magnetic fields.
\end{abstract}
\maketitle
\paragraph{Introduction.---}
Vortex beams, characterized by their quantized orbital angular momentum (OAM) along the axis of propagation and helical wavefronts, represent a powerful tool endowed with an additional degree of freedom for exploring new physics~\cite{Bliokh_20170524,Ivanov_202211__}. Researches on vortex beams have advanced across multiple disciplines, including particle physics~\cite{Sarenac_20220918,Ivanov_20231204,Karlovets_20230506,Zou_20230515}, optics~\cite{Bliokh_20150826,Bliokh_20210616,Shen_20191002,Jentschura_20110105,Wei_20200827}, nuclear physics~\cite{Lu_20231114,Wu_20220422,Xu_202405__}, atomic and molecular physics~\cite{Luski_20210901,Maslennikov_20240502,Quinteiro_20171220}, condensed matter physics~\cite{Mendis_202209__,Rosen_20220825,Silenko_20240103,Rondón_20191204,Mizushima_20231004}, and astrophysics~\cite{Tamburini_20210721}, among others. Within this broad range, vortex electrons have been extensively investigated~\cite{Uchida_20100401,Verbeeck_20100901,McMorran_20110114,Bliokh_20111017,Bialynicki-Birula_20170313,Barnett_20170313,Saitoh_20130814,Grillo_20170925}. Specifically, consistent efforts are directed towards understanding their behavior in magnetic fields ~\cite{Rajabi_20170612,Kruining_20170718,Zou_20210106,Greenshields_20121025,Greenshields_20150909,Bliokh_20120916,Guzzinati_20130225,Schattschneider_20140808,Schachinger_201509__,Gallatin_20120705,Greenshields_20141209,Karlovets_20210330} and one widely discussed topic is the rotational dynamics of EVBs~\cite{Guzzinati_20130225,Bliokh_20120916,Schattschneider_20140808,Schachinger_201509__}.

\begin{figure}[h!]
	\centering
	\includegraphics[width=0.55\linewidth]{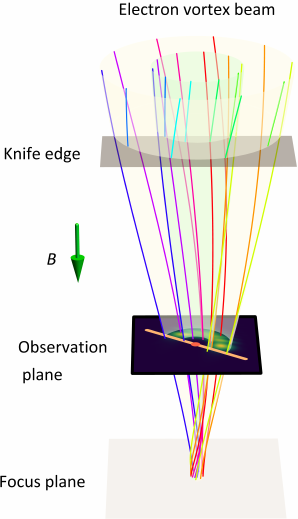}
	\caption{The Bohmian trajectories of electron vortex beam in uniform magnetic fields. A knife-edge cuts the vortex beam, producing a truncated pattern that rotates according to the Bohmian trajectories. Trajectories with the same initial radius are distinguished by hue color mapping based on different initial angles. The expectation value of the rotation angle is highlighted by a light orange line.}
	\label{fig:bohmiantraj}
\end{figure}

Understanding the rotational dynamics of EVBs in magnetic fields, particularly in the context of transmission electron microscopy (TEM)~\cite{Bliokh_20170524,Guzzinati_20130225, Barrows_20221212,Schattschneider_20140808,Schachinger_201509__}, presents a complex challenge both experimentally and theoretically. To approach this problem, it is helpful to first recall the classical concept of Larmor rotation in TEM~\cite{Kohl_20080828} before addressing quantum aspects. When an electron beam passes through a magnetic lens—standard in TEM setups—the resulting image is rotated around the symmetric axis of the lens system by an angle that depends on the magnetic field strength. This image rotation represents a coherent rotation of the electron beam as a whole around the fixed symmetry axis of the magnetic lens. In contrast, the cyclotron motion of individual electrons occurs locally, around the center of curvature of each electron's helical trajectory.
	
In quantum mechanics, a richer picture of rotational dynamics emerges—one that goes beyond classical trajectories and uncovers intrinsic OAM carried by the wavefunction itself. Even when electrons propagate along the symmetric axis of the magnetic lens with negligible transverse velocity—thus precluding classical rotation—quantum mechanics still permits complex internal rotation arising from the wavefunction's structured phase and probability currents. A prime example of such intrinsic rotational behavior is found in the eigenstates of electrons in a uniform magnetic field, commonly referred to as Landau states~\cite{Bliokh_20120916,Schattschneider_20140808,Schachinger_201509__}. In particular, these spatially structured Landau modes feature circulating probability currents and phase singularities that give rise to internal rotation within the beam. This internal dynamics is elegantly illustrated by Bohmian trajectories, which represent the streamlines of the probability current, tracing spiral paths that demonstrate a fundamental rotational characteristic inherent to the quantum state, as shown in Fig.~\ref{fig:bohmiantraj}.

Although this intrinsic rotation is always present, it becomes experimentally observable only when the beam's cylindrical symmetry is broken—for example, by partially blocking the beam. In such cases, the internal angular evolution manifests as a measurable rotation of the beam's truncated intensity pattern in the transverse plane. This observable effect is often taken as the signature of the quantum beam's rotational dynamics and provides insight into the internal structure of vortex beams in magnetic fields.

Recent studies reveal seemingly distinct regimes of rotational behavior: in a specific propagation region, the beam's angular frequencies can be approximated by three fixed frequencies predicted by the Landau states~\cite{Schattschneider_20140808,Schachinger_201509__}; near the focus, a rapid rotation appears~\cite{Schachinger_201509__,Guzzinati_20130225}, akin to the free-space Gouy rotation; and in the far field, the rotation approaches the classical Larmor frequency~\cite{Schachinger_201509__,Guzzinati_20130225}. Remarkably, these rotational regimes have all been observed within a single beam at varying propagation distances~\cite{Schachinger_201509__}, but lacks a unified description. This motivates a re-evaluation of the rotational dynamics, which may contribute to spatially resolved measurements of longitudinal magnetic fields~\cite{Greenshields_20121025} and manipulations of nanoparticles with EVBs~\cite{Verbeeck_20121126} in magnetic fields.

The relativistic paraxial equation in magnetic fields for electron beams can be derived from the Foldy-Wouthuysen (FW) Hamiltonian~\cite{Zou_20210106,Silenko_20220622}. The FW representation \cite{Foldy_19500401} establishes a Schrödinger picture of relativistic quantum mechanics \cite{Foldy_19500401,Zou_20200326_2}. The exact solution of this equation proves to be a promising candidate for investigating EVBs in magnetic fields~\cite{Zou_20210106,Sizykh_20240413}, referring hereafter as ``paraxial Landau modes". The Landau states, commonly used to characterize electrons in uniform magnetic fields, emerge as special cases when the beam waist of paraxial Landau modes equals a magnetic length parameter~\cite{Zou_20200326}. Furthermore, as magnetic fields approach zero, the paraxial Landau modes transform into the free-space Laguerre-Gaussian (LG) beams~\cite{Bliokh_20170524}.

The Gouy phase is a fundamental phase anomaly of propagating waves~\cite{Siegman_19861017,Feng_20010415}. While extensively studied in optics~\cite{Hiekkamaki_20221006,Gu_20180307,Kawase_20080729,Zhong_20210524,Liebmann_201710__,Yavorsky_20230418}, research on the Gouy phase of matter waves~\cite{Guzzinati_20130225,daPaz_20111208,Brennecke_20200415,Ducharme_20150828,daPaz_20161212,Marinho_20240528} and its impact on quantum state evolution remains limited. In free space, the Gouy phase is associated with the rotational dynamics of vortex beams~\cite{Hamazaki_200609__,Baumann_200906__,Guzzinati_20130225}. However, the original form of Gouy rotation inadequately describes rotation of EVBs within magnetic fields. The Gouy phase of the paraxial Landau modes takes on a distinct form compared to the familiar Gouy phase of free-space LG beams. This leads to a generalized Gouy rotation associated with paraxial Landau modes, which proves useful in describing rotational dynamics of EVBs in uniform magnetic fields.
 
In this work, we unify the different rotational regimes of EVBs in uniform magnetic fields by introducing and deriving the generalized Gouy rotation frequency. This quantity links the phase structure of EVB wavefunctions in magnetic fields directly to observable rotational dynamics. It reveals reversals in the rotation direction for EVBs with $\ell<0$, a feature that has not been coherently incorporated into existing theoretical models. The rotation angles of EVBs are theoretically calculated based on paraxial Landau modes, demonstrating consistency with previously published experimental data in~\cite{Schachinger_201509__}. To further validate our findings, we conduct numerical simulations using the Chebyshev method. In the Discussions, we distinguish our analytical single-mode analysis from an alternative effective method proposed in~\cite{Schachinger_201509__}, which involves representing beams in magnetic fields as superpositions of free LG beams.

\paragraph{Paraxial Landau modes.---}
The FW Hamiltonian is $\hat{H}_{\textrm{FW}}=\beta\sqrt{m^2c^4+\boldsymbol{\hat{\pi}}^2c^2-e\hbar c^2\boldsymbol{\Sigma}\cdot \boldsymbol{B}}$~\cite{Case_19540901,Zou_20210106,Silenko_20220622}, where $\boldsymbol{\hat{\pi}}=\boldsymbol{\hat{p}}-e\boldsymbol{\hat{A}}$ represents the kinetic momentum, $\boldsymbol{\hat{A}}=\boldsymbol{\hat{B}}\times\boldsymbol{\hat{r}}/2$ is the vector potential in symmetric gauge, $\beta$ and $\boldsymbol{\Sigma}$ are the Dirac matrices. The $z$ axis of cylindrical coordinates $(r,\phi,z)$ is directed along magnetic field, $\boldsymbol{B}=B\bar{\boldsymbol{z}}$. The electron charge is $e=-|e|$.  The spin term $s_zB$ can be disregarded because it can be eliminated together with $m^2$ after the zero energy level shift~\cite{Zou_20210106}. Using the paraxial approximation $\boldsymbol{\hat{\pi}}^2\approx p\hat{p}_z+\boldsymbol{\hat{\pi}}_{\perp}^2/2
$ together with the ansatz $\psi=e^{ikz}\Psi$, one can obtain the paraxial equation for EVBs in uniform magnetic fields~\cite{Zou_20210106}:
\begin{equation}\label{eq:par_eq}
	\left[2i\hbar^2 k\frac{\partial}{\partial z}+\hbar^2\nabla_{\perp}^2-i\hbar eB\frac{\partial}{\partial \phi}-\frac{1}{4}e^2B^2r^2\right]\Psi=0,
\end{equation}
where $k$ is the wave number related to momentum by the de Brogile relation $p=\hbar k$, and $p$ satisfies the relativistic dispersion relation $E^2=m^2c^4+p^2c^2$. $\nabla_{\perp}^2$ is the transverse Laplace operator in cylindrical coordinates. The \emph{exact} form of Eq.~(\ref{eq:par_eq}) has been obtained in \cite{Silenko_20220622}.

The exact solution of the paraxial equation~(\ref{eq:par_eq}), given in earlier studies ~\cite{Zou_20210106,Melkani_20210716,Sizykh_20240401,Sizykh_20240413}, is
\begin{subequations}\label{eq:par_sol}
	\begin{align}
		\Psi_{n\ell}&=Ae^{i\left(\ell\phi+\frac{kr^2}{2R(z)}-\Phi_G(z)\right)},\\
		A&=\frac{C_{n\ell}}{w(z)}\left(\frac{\sqrt{2}r}{w(z)}\right)^{|\ell|}L_n^{|\ell|}\left(\frac{2r^2}{w(z)^2}\right)e^{-\frac{r^2}{w(z)^2}},\\
		C_{n\ell}&=\sqrt{\frac{2n!}{\pi(n+|\ell|)!}}.
	\end{align}    
\end{subequations}
$A$ defines the real amplitude. $L_n^{|\ell|}$ is the generalized Laguerre polynomial, with $n=0,1,2,\cdots$ the radial quantum number and $\ell=0,\pm 1,\pm 2,\cdots$ the topological charge. $w(z)$ is the beam width, $R(z)$ is the radius of curvature of the wavefront, and $\Phi_G(z)$ is the Gouy phase~\footnote{Note that these are derived under the assumption of $w'(0)=0$. This assumption is specific to the experimental setup where $z=0$ corresponds to the focal point.}:
\begin{subequations}\label{eq:par_parms}
	\begin{align}
		w(z)&=w_0 \sqrt{\cos ^2 \frac{z}{z_m}+\frac{z_m^2}{z_R^2} \sin ^2 \frac{z}{z_m}}\label{eq:beamwidth},\\
		R(z)&=k w_m^2 \frac{\cos ^2 \frac{z}{z_m}+\frac{z_m^2}{z_R^2} \sin ^2 \frac{z}{z_m}}{\left(\frac{z_m^2}{z_R^2}-1\right) \sin \frac{2 z}{z_m}},\\
		\Phi_G(z)&=N\arctan \left(\frac{z_m}{z_R} \tan \frac{z}{z_m}\right)+\ell \frac{z}{z_m}\label{eq:Gouy},
	\end{align}
\end{subequations}
where $N=2n+|\ell|+1$, $w_0$ is the beam waist, $z_R=kw_0^2/2$ is the Rayleigh distance, $w_m=2\sqrt{\hbar/|e|B}$ is a magnetic length parameter, and $z_m=k w_m^2/2$ is related to the Larmor distance through $z_L=2\pi z_m$~\cite{Schattschneider_20140808}. When taking the limit $B\to 0$, the parameter functions in Eq.~(\ref{eq:par_parms}) become that of free-space LG beams~\cite{Zou_20210106}. Once the condition $w_m=w_0$ is satisfied, Landau states are recovered~\cite{Zou_20200326}. The beam width given by Eq.~(\ref{eq:beamwidth}) oscillates with a spatial period of $\pi z_m$, similar to the classical and well-known periodic oscillations of electron trajectories inside magnetic fields. While the latter results from the Lorentz force and the electron's classical motion, the oscillations of the beam width arise from the beam's quantum wave nature. It is worth noting that inside the typical Glaser lens field of a TEM objective lens the classical trajectory can oscillate with varying amplitudes and non-strict periodicity.
\paragraph{Rotational Dynamics.---}

\begin{figure}
	\centering
	\includegraphics[width=\linewidth]{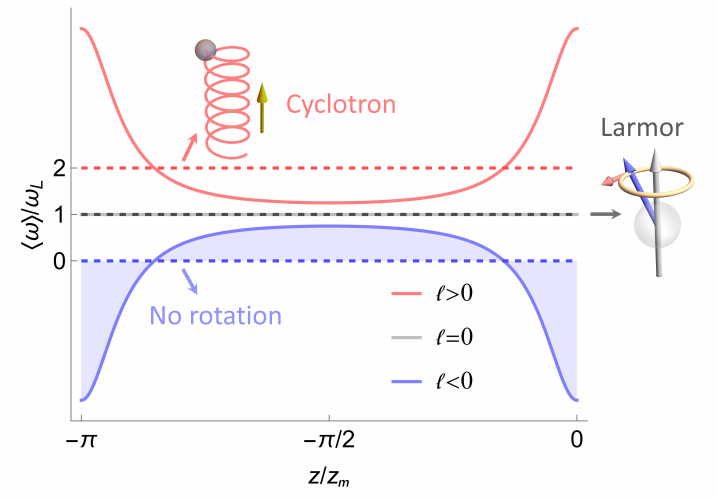}
	\caption{Normalized angular frequency $\langle \omega\rangle/\omega_L$ as a function of $z/z_m$. The solid curve is obtained from the paraxial Landau modes by substituting Eq.~(\ref{eq:beamwidth}) into Eq.~(\ref{eq:ro_angfre}), while the dashed lines indicate the quantized angular frequencies of the Landau states~\cite{Schattschneider_20140808}. For $\ell<0$, the shaded region highlights the sign reversal of $\langle\omega\rangle$ across the $\langle\omega\rangle = 0$ axis. This plot serves as an illustrative example, where we set $z_m = 4z_R$ and display one representative periodic interval from $-\pi z_m$ to $0$. The two insets provide classical schematic views of cyclotron and Larmor frequencies for reference.}
	\label{fig:angularfrequency}
\end{figure}

The rotation of EVBs in a magnetic field is closely linked to Bohmian trajectories \cite{Bliokh_20170524,Schattschneider_20140808,Schachinger_201509__}. These trajectories illustrate the spiralling motion of electrons around the direction of magnetic fields.
When considering the presence of a vector potential $\boldsymbol{A}$, the gauge-invariant probability current is $\boldsymbol{j}=\left[\hbar \text{Im}\left(\psi^{*}\boldsymbol{\nabla}\psi\right)-e\boldsymbol{A}|\psi|^2\right]/m$~\cite{Bliokh_20170524,CohenTannoudji_20191216}. The angular frequency of electron is related to the current through $\omega(r)=v_{\phi}(r)/r=\hbar\ell/m r^2+\omega_L$ with the local Bohmian velocity given by $\boldsymbol{v} = \boldsymbol{j} / \rho$ where $\rho = |\psi|^2$ and $\omega_L = |e|B / 2m$ is the Larmor frequency. Its expectation value turns out to be
\begin{equation}\label{eq:ro_angfre}
	\left\langle \omega\right\rangle(z)=\omega_L\left(\text{sgn}(\ell)\frac{w_m^2}{w(z)^2}+1\right),
\end{equation}
where we have used the fact that for LG beam given in the form of Eq.~(\ref{eq:par_sol}) and $\ell\not=0$, $\left\langle r^{-2}\right\rangle=2/|\ell|w(z)^2$~\cite{Schachinger_201509__,Zwillinger_20140918}. For Landau states, $w(z)=w_m$ and we recover the quantized rotation of Landau electrons: $\left\langle \omega\right\rangle(z)=\omega_L\left(\text{sgn}(\ell)+1\right)$~\cite{Schattschneider_20140808,Bliokh_20120916}. Depending on the sign of topological charge $\ell$, it can take values of cyclotron frequency $\omega_c=2\omega_L$, Larmor frequency $\omega_L$, and zero, as illustrated in Fig.~\ref{fig:angularfrequency}. A notable observation is that, unlike the zero angular frequency predicted by conventional Landau states for $\ell<0$, the generalized Gouy rotation can exhibit reversals in the rotation direction along propagation. This behavior is also absent in free-space Gouy rotation and highlights a distinctive effect induced by the magnetic field.

Changing the time increment to experimentally accessible $z$-shift $\left\langle\omega\right\rangle=d\left\langle\phi\right\rangle/dt= vd\left\langle\phi\right\rangle/dz$, we can then calculate the Bohmian rotation angle through $\left\langle\phi\right\rangle=\int\left\langle\omega\right\rangle dz/v$:
\begin{equation}\label{eq:ro_ang}
	\left\langle\phi\right\rangle=\text{sgn}~(\ell)\arctan\left(\frac{z_m}{z_R}\tan\left(\frac{z}{z_m}\right)\right)+\frac{z}{z_m}.
\end{equation}
Here the integration constant is chosen to be zero and $v=\hbar k/m$ is assumed to be uniform. The relationship between the Bohmian rotation angle and the Gouy phase can be interpreted by Coriolis (or rotational-Doppler) coupling between the \emph{intrinsic} OAM carried by the EVBs and the Bohmian coordinate-frame rotations~\cite{Bliokh_20080716}: \(\Phi_{\rm Coriolis} = -\int \mathbf{J} \cdot \boldsymbol{\Omega} \, dt,\) with \(\mathbf{J} = \ell \hat{\mathbf{e}}_z\), and \(\boldsymbol{\Omega} = \langle \omega \rangle \hat{\mathbf{e}}_z\).

The formula~(\ref{eq:ro_ang}) for rotation angles can be related to the experimental data in \cite{Schachinger_201509__}. In the experiment, an incident electron beam was initially transformed into EVBs using a holographic fork mask. The EVBs then encountered a longitudinal magnetic field produced by the objective lens inside the TEM. The EVBs continued to propagate until they reach a knife-edge (KE). Subsequently, the cut EVBs proceeded along the column, ultimately reaching the observation plane positioned above focal plane, i.e. the defocus position $z_{df}$. The rotational dynamics of the EVBs are studied by adjusting the position of the KE $z_k$, resulting in variations in azimuthal angles of the half-moon-like patterns on the observation plane:
\begin{equation}\label{eq:ang_diff}
\begin{aligned}
\Delta\langle\phi\rangle&=\text{sgn}~(\ell)\left[\arctan\left(\frac{z_m}{z_R}\tan\left(\frac{z_k}{z_m}\right)\right)\right.\\
&~~\left.-\arctan\left(\frac{z_m}{z_R}\tan\left(\frac{z_{df}}{z_m}\right)\right)\right]+\frac{z_k-z_{df}}{z_m}.
\end{aligned}
\end{equation}
Note that the arc tangent term in Eq.~(\ref{eq:ang_diff}) is understood hereafter as $\left(\arctan \left(\frac{z_m}{z_R} \tan \frac{z}{z_m}\right)+\pi \left\lfloor\frac{z}{\pi z_m}+\frac{1}{2}\right\rfloor\right)$ to ensure its continuity over $(-\infty,\infty)$, where $\lfloor\cdot\rfloor$ denotes the floor function~\cite{Sizykh_20240413}. For clarity, all patterns are observed at the same fixed defocus plane $z_{df}$; each pattern is generated in a separate trial by truncating the beam at a different knife-edge position $z_k$, and is labeled accordingly by that $z_k$.

As a heuristic argument to support the use of Eqs.~(\ref{eq:ro_ang}--\ref{eq:ang_diff}) for describing the rotation of truncated EVBs, we refer to Fig.~\ref{fig:bohmiantraj}, where it is assumed that the Bohmian trajectories in the untruncated region remain largely unaffected by the truncation. A more rigorous argument relies on the assumptions that truncation preserves the radial profile of the original vortex beam~\cite{Schachinger_201509__} and conserves the expectation value of OAM~\cite{Guzzinati_20130225,Schachinger_201509__}.  Suppose the uncut vortex $\Psi \propto e^{i\ell\phi}$ is modified only in its angular dependence by a factor $f(\phi)$, such that the truncated wavefunction becomes $\Phi \propto f(\phi)e^{i\ell\phi}$. The angular frequency shift is then given by $\Delta \omega\propto \text{Im}(f^*\partial_{\phi}f)/|f|^2$, whose expectation value is $\Delta \langle\omega\rangle\propto \int_{0}^{2\pi}d\phi~\text{Im}(f^*\partial_{\phi}f)$. Since $\phi=0$ and $\phi=2\pi$ represent the same physical point, the function $f(\phi)$ is $2\pi$-periodic and admits a Fourier expansion: $f(\phi)=\sum_{s}c_{s}e^{is\phi}$. It follows that
	\begin{equation}
		\Delta \langle\omega\rangle\propto \text{Re}\left(\int_{0}^{2\pi}d\phi~\sum_{s',s}sc_{s'}^{*}c_{s}e^{i(s-s')\phi}\right)=\sum_{s}s|c_s|^2.
	\end{equation}
	Meanwhile, the OAM expectation value is
	\[
	\begin{aligned}
		\langle \hat{L}_z\rangle_{\Phi}&=\int_{0}^{2\pi}d\phi~\sum_{s',s}\hbar(\ell+s)c_{s'}^*c_se^{i(s-s')\phi}\\
		&=\hbar\left(\ell+\sum_{s}s|c_s|^2\right),
	\end{aligned}
	\]
	where we have used the normalization $\sum_{s}|c_{s}|^2=1$. Thus, ensuring $\Delta \langle \omega \rangle = 0$ is equivalent to requiring that $\langle \hat{L}_z \rangle_{\Phi} = \hbar \ell$, i.e., $\sum_s s|c_s|^2 = 0$. In this case, the truncation $f(\phi)$ may broaden the OAM distribution but does not shift its mean. Both existing experimental results and our subsequent simulations are consistent with these assumptions, confirming that the observed rotation of the truncated beam reliably reflects the intrinsic rotational dynamics of the vortex beam.

\begin{figure}
	\centering
	\includegraphics[width=\linewidth]{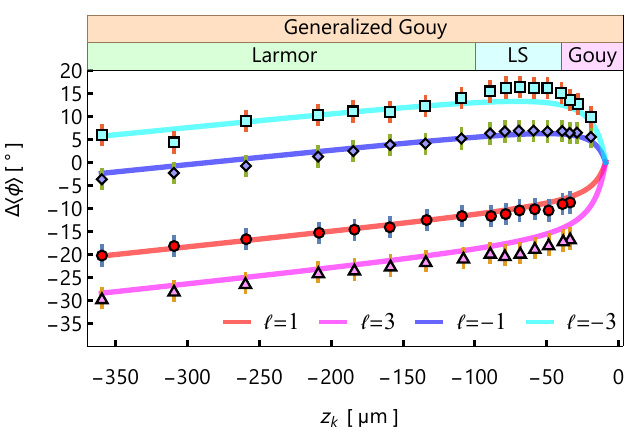}
	\caption{The calculated rotation angles $\Delta\langle \phi\rangle$ as functions of knife-edge positions $z_k$. The upper annotations reproduce the region divisions from Ref.~\cite{Schachinger_201509__} (Larmor, LS, and Gouy) and introduce the ``Generalized Gouy'', a new perspective that brings these regions together under a unified framework. The cyan squares, the blue rhombus, the red circles, and the magenta triangles represent the experimental data, taken from~\cite{Schachinger_201509__}, for EVBs with $n=0$ and $\ell=-3,-1,1,3$, respectively. The solid lines represent our theoretical result in Eq.~(\ref{eq:ang_diff}), and although the curves are similar to those from the effective model used in Ref.~\cite{Schachinger_201509__} within the limited experimental parameter ranges, they are independently derived and reflect the current theoretical framework. The Rayleigh distances used in the experiment are $z_R=1.46$~\textmu m for $|\ell|=1$ and $z_R=2.84$~\textmu m for $|\ell|=3$; the observation plane is shifted from the focus plane by a distance $z_{df}=-9$~\textmu m. Note that $\Delta \langle \phi\rangle$ is plotted in degree, requiring an additional factor of $180/\pi$ in Eq.~(\ref{eq:ang_diff}).}
	\label{fig:expdataplot}
\end{figure}

In the experiment~\cite{Schachinger_201509__}, the magnetic field used is $B=1.9~\textrm{T}$ for a standard TEM objective lens. Under the working energy of $200~\textrm{keV}$, the electron travels at a relativistic speed $v\simeq 0.7c$, where $c$ is the speed of light. The relativistic mass of the electron is $m=\gamma m_0$, where $m_0$ denotes the rest electron mass. Utilizing $w_m=2\sqrt{\hbar/|e|B}$ and the wave number $k=mv/\hbar$, we can determine the characteristic distance: $z_m=kw_m^2/2\simeq$ 1759~\textmu m. 

The theoretical analysis based on Eq.~(\ref{eq:ang_diff}) is plotted in Fig.~\ref{fig:expdataplot} for EVBs with different topological charges $\ell=\pm1,\pm3$, which are consistent with the experimental data, taken from \cite{Schachinger_201509__}. In the experimental conditions utilized in \cite{Schachinger_201509__}, the Gouy rotation exhibits notably rapid behavior near the focal point. Importantly, in the region ranging from $-80$~\textmu m to $-30$~\textmu m, we have the approximate Landau states characterized by $w(z)\simeq w_m$, in agreement with the experiment in \cite{Schattschneider_20140808}. The experiment in \cite{Guzzinati_20130225} made separate observations of the Gouy rotation and Larmor rotation in magnetic fields. In our context they can be considered as components of the generalized Gouy rotation that dominate across different ranges. As shown in Fig.~\ref{fig:expdataplot}, EVBs with $\ell=-1$ and $\ell=-3$ exhibit a reversal in the rotation direction, supporting the characteristic behavior of generalized Gouy rotation for $\ell<0$.

\paragraph{Simulations.---}

To further validate our theoretical analysis and enable subsequent discussions, we perform numerical simulations using the paraxial equation~(\ref{eq:par_eq}) as the governing equation. The simulations are initialized with the truncated paraxial Landau modes, described by
\begin{equation}\label{eq:initialstate}
	\Phi(\boldsymbol{r})|_{z=z_k}=\left\{
	\begin{aligned}
		\Psi_{n\ell}(\boldsymbol{r})|_{z=z_k}&,~\phi\in (0,\pi)\\
		0\qquad&,~\phi\in (\pi,2\pi).
	\end{aligned}
	\right.
\end{equation}
It is worth noting that the truncated beam $\Phi$ represents a superposition of numerous paraxial Landau modes which interfere and result in complex diffraction deformations~\cite{Guzzinati_20130225}. Its OAM expectation value proves to be approximately conserved, $\langle \hat{L}_z\rangle\approx \hbar \ell$~\cite{Guzzinati_20130225,Schachinger_201509__}. 

The numerical method used here is the Chebyshev method. Although it is originally applied to solve the time-dependent Schr{\"o}dinger equation (TDSE)~\cite{Greenshields_20150909,Izaac_20190215,Leforestier_199105__}, we notice that Eq.~(\ref{eq:par_eq}) can be written as a TDSE-like form:
\begin{equation}\label{eq:par_eq_TDSE}
	i\hbar\frac{\partial}{\partial z}\Phi=\left[-\frac{\hbar}{2k}\nabla_{\perp}^2+i\frac{eB}{2k}\frac{\partial}{\partial \phi}+\frac{e^2B^2r^2}{8\hbar k}\right]\Phi.
\end{equation}
\begin{figure}
	\centering
	\includegraphics[width=0.8\linewidth]{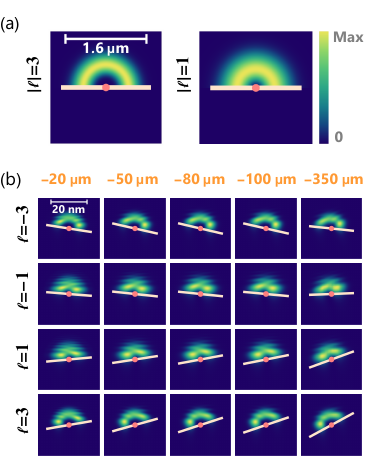}
	\caption{(a) The initial intensity profiles of $\Phi$ on the KE plane at $z_k=-350$~\textmu m as reference. (b) The simulated intensity profiles of $\Phi$ on the observation plane for EVBs with radial index $n=0$ and topological charge $\ell=-3,-1,1,3$ respectively, at five distinct cutting positions: $z_k=-20$~\textmu m, $-50$~\textmu m, $-80$~\textmu m, $-100$~\textmu m, $-350$~\textmu m. The light orange lines overlaid on the intensity profiles indicate the opposite of theoretically calculated angles described in Eq.~(\ref{eq:ang_diff}) since we are comparing the intensity profiles at the observation plane with the KE plane. Remarkably, these lines are in accordance with the observed rotation angles of the intensity profiles.}
	\label{fig:sim}
\end{figure}

\begin{figure}
	\centering
	\includegraphics[width=\linewidth]{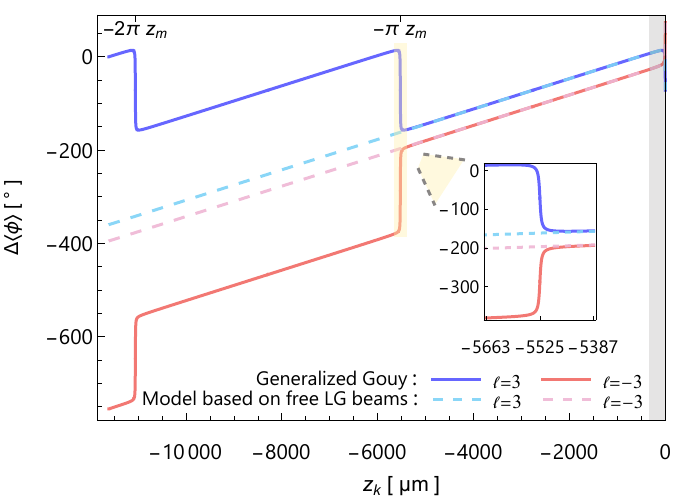}
	\caption{Rotation angles $\Delta\langle \phi\rangle$ as functions of $z_k$ for generalized Gouy rotation and model based on free LG beams. We use the Rayleigh distance $z_R=2.84$~\textmu m for $n=0,|\ell|=3$ as in the experiment in~\cite{Schachinger_201509__} and extend the $z$-shift range down to $-2\pi z_m$ for differentiation between the two approaches. The gray band represents the accessed $z$-shift range in the experiment. The region near the focus at $-\pi z_m$ is magnified to illustrate the fast yet continuous Gouy rotation.}
	\label{fig:discussion1}
\end{figure}

\begin{figure*}
	\centering
	\includegraphics[width=0.75\linewidth]{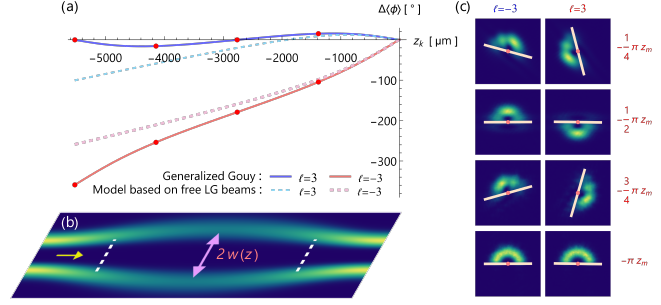}
	\caption{(a)~Rotation angles $\Delta\langle \phi\rangle$ as functions of $z_k$ for generalized Gouy rotation and model based on free LG beams. The Rayleigh distance used here is $z_R=1000$~\textmu m, in contrast to the experimental value of $2.84$~\textmu m for $n=0,|\ell|=3$. The red points indicate the locations where simulations are conducted. (b)~The longitudinal slice of the beam intensity profile illustrates the change in beam width $w(z)$, indicating beam undergoes both divergence and convergence along its propagation. The white dashed lines mark the magnetic length $w_m$. The yellow arrow indicates the direction of beam propagation. (c)~The simulated intensity profiles for $n=0,|\ell|=3$ at different cutting positions $z_k=-\pi z_m/4,-\pi z_m/2,-3\pi z_m/4,-\pi z_m$. The light orange lines represent the theoretical predictions based on Eq.~(\ref{eq:ang_diff}), which are consistent with simulations. The scale bar is identical to that in Fig.\ref{fig:sim}~(a), and the size of the field of view is the same as that of Fig.\ref{fig:sim}~(b).}
	\label{fig:discussion2}
\end{figure*}

The transverse spatial extent of the simulated wavefunction decreases by a ratio of 6400 from the smallest $z$-shift to the observation plane, as shown in Fig.~\ref{fig:sim}. To ensure stability and precision in the simulation, a fine spatial grid is necessary. Acting as a global approximation technique, the Chebyshev method facilitates the direct calculation of the final states once the Hamiltonian and initial states of the system are provided. This is achieved by expanding the unitary propagation operator as a series of Chebyshev polynomials, as discussed in~\cite{Izaac_20190215,Greenshields_20150909,Leforestier_199105__}. Additional numerical details are provided in the Supplemental Material~\footnote{See Supplemental Material [url], which includes Refs.~\cite{Izaac_20190215,Greenshields_20150909,Leforestier_199105__,Greenshields_20150909}, for simulation details, parameter settings, and computational cost.}.

The simulated results are shown in Fig.~\ref{fig:sim}. The rotations of semi-annular intensity profiles can be seen to match the dashed lines representing our theoretical analysis in Eq.~(\ref{eq:ang_diff}). Specifically, at $z_k=-20$~\textmu m and $-50$~\textmu m, the intensity profiles exhibit rapid Gouy rotation. Subsequently, from $-50$~\textmu m to $-80$~\textmu m, the intensity profiles  for $\ell<0$ undergo minimal rotation, indicative of the zero frequency associated with the Landau states. Noteworthy observations emerge from $-100$~\textmu m to $-350$~\textmu m. EVBs with positive $\ell$ values exhibit consistent rotation in one direction, while those with negative $\ell$ values demonstrate a reversal of rotation, as can be observed in Fig.~\ref{fig:sim}~(b).
\paragraph{Discussions.---}
Our method using paraxial Landau modes is different from that used in \cite{Schachinger_201509__}. The latter involves decomposing the wavefunction in magnetic fields into free LG beams and finds that for converging EVBs in magnetic fields with vortex order $\ell$, free LG beams of the same order $\ell$ provide a viable approximation. By substituting the beam width $w(z)=w_0\sqrt{1+z^2/z_R^2}$ of free LG beams into Eq.~(\ref{eq:ro_angfre}), one can derive a formula for the rotation angles:
\begin{equation}\label{eq:ang_diff_free}
	\begin{aligned}
		\Delta\langle\phi\rangle&=\text{sgn}~(\ell)\left[\arctan\left(\frac{z_k}{z_R}\right)-\arctan\left(\frac{z_{df}}{z_R}\right)\right]\\
		&~~+\frac{z_k-z_{df}}{z_m}.
	\end{aligned}
\end{equation}
In the range of accessible $z$-shifts and under the parameters employed in their experiment~\cite{Schachinger_201509__}, it shows effectiveness in describing converging EVBs. However, this approach combines characteristics of the rapid, free-space Gouy-like rotation near the focus with the Larmor frequency induced by the magnetic field, yet fails to provide a coherent physical description. In contrast, our formulation unifies these effects, clearly demonstrating how the beam undergoes a rotation driven by the Gouy phase of paraxial Landau modes, with the magnetic field's influence inherently embedded in the solution.

We take into account the oscillating behavior of beam width induced by magnetic fields. The range of $z$-shifts can be extended such that the beam demonstrates both divergent and convergent behavior and the result is shown in Fig.~\ref{fig:discussion1}. To further distinguish between the two approaches and facilitate simulations, we opt for a Rayleigh distance of $z_R=1000$~\textmu m. This amounts to choosing a beam waist of approximately $20~\textrm{nm}$ rather than the $1~\textrm{nm}$ used in the experiment. The results are shown in Fig.~\ref{fig:discussion2} and it should be noticed that for $z_k\le -\pi z_m/2$ the beam undergoes divergence and the difference becomes apparent.
	
Further experimental verifications could involve employing a low-energy TEM~\cite{Sasaki_201410__}, or increasing the magnetic field. Both strategies aim to reduce $z_m=2m v/|e|B$ such that the critical position $-\pi z_m$ lies within the accessible region of the experiment, as depicted in Fig.~\ref{fig:discussion1}. Importantly, while the magnetic field is considered uniform within the studied propagation range, the experiments were conducted in the approximately bell-shaped Glaser field~\cite{Melkani_20210716,Loffler_2020____,Khan_202103__} of a TEM objective lens. Extensions to Glaser fields remain a subject for future exploration.

\paragraph{Conclusions.---}

The generalized Gouy rotation is associated with the Gouy phase of the paraxial Landau modes, which extend Landau states by allowing their transverse structure to evolve during propagation. Unlike the free-space Gouy rotation, which is characterized by rapid image rotation near the focus and remains nearly constant beyond the Rayleigh distance, the generalized version exhibits distinct behaviors throughout the entire beam propagation. Importantly, it enables novel predictions of angular frequencies that go beyond the three fixed values allowed by conventional Landau states. In particular, it can account for reversals in the rotation direction of electron vortex beams with negative topological charge.

Our model is consistent with both experimental observations and numerical simulations. This agreement suggest that paraxial Landau modes may arise naturally in realistic settings with quasi-uniform magnetic fields. We also identify parameter regimes where the effects of generalized Gouy rotation are expected to become more pronounced, suggesting viable directions for future experimental verification. Moreover, our findings are not limited to the context of electron microscopy and, in principle, apply to any system involving electron vortex beams in uniform magnetic fields. Extensions to nonuniform configurations, such as Glaser fields, represent a natural direction for future work.

\paragraph{Acknowledgements.---}
We thank I. P. Ivanov for valuable comments. The work was supported by the National Key R\&D Program of China No. 2024YFE0109802, and the National Natural Science Foundation of China (grants No.~12175320 and No.~12375084) and Guangdong Basic and Applied Basic Research Foundation (grants No.~2022A1515010280 and No.~2022B1515120027).

\end{document}